\documentclass[journal, 9pt]{IEEEtran}

\def\TIFS{0}

\usepackage{amsmath}
\usepackage{amssymb}
\usepackage{amsthm}
\usepackage{xcolor}

\definecolor{codegreen}{rgb}{0,0.6,0}
\definecolor{codegray}{rgb}{0.5,0.5,0.5}
\definecolor{codepurple}{rgb}{0.58,0,0.82}

\usepackage{url}

\newtheorem{lemma}{Lemma}

\begin{document}

\date{}

\title{Comments on ``A Privacy-Preserving Online Ride-Hailing System Without Involving a Third Trusted Server''}

\ifnum\TIFS=1
\author{
Srinivas Vivek\orcidlink{0000-0002-8426-0859}
\thanks{S. Vivek is with the International
Institute of Information Technology Bangalore, Bengaluru 560 100, India. Email: srinivas.vivek@iiitb.ac.in.   
}
}
\else
\author{
Srinivas Vivek
\thanks{ The ORCID ID of S. Vivek is {0000-0002-8426-0859}. S. Vivek is with the International
Institute of Information Technology Bangalore, Bengaluru 560 100, India. Email: srinivas.vivek@iiitb.ac.in.   
}
}
\fi

\ifnum\TIFS=1
\markboth{IEEE Transactions on Information Forensics and Security}%
{S. Vivek: Comments on ``A Privacy-Preserving Online Ride-Hailing System \ldots''}
\else
\markboth{}%
{S. Vivek: Comments on ``A Privacy-Preserving Online Ride-Hailing System \ldots''}
\fi

\maketitle

\begin{abstract}
Recently, Xie et al. (IEEE Transactions on Information Forensics and Security, vol. 16, pp. 3068-3081, 
2021) proposed a privacy-preserving Online Ride-Hailing (ORH) protocol that does not 
make use of a trusted third-party server. The primary
goal of such privacy-preserving ORH protocols is to ensure the privacy of riders' and
drivers' location data w.r.t. the ORH Service Provider (SP). In this note, 
we demonstrate a passive attack by the SP in the protocol of Xie et al. that enables it to
\textit{completely} recover the location of the rider as well as that of the responding drivers 
in each and every ride request query.
 
\end{abstract}
%
\begin{IEEEkeywords}
Ride-Hailing Services, Privacy, Attack.
\end{IEEEkeywords}

%

\section{Introduction}

Online Ride-Hailing (ORH) services offer the convenience of ride matching at the tip of fingers 
and have become very popular in
recent times. The privacy-preserving ORH by Xie et al. \cite{XieGJ21} is one of the many recent
ORH proposals that is designed to offer 
accurate ride matching service by the ORH Service Provider (SP) without the SP actually learning information about the 
locations of riders and drivers. Further, their protocol does not make use of a trusted third-party during
the ride-matching process. In order to ensure an accurate and privacy-friendly match, the locations are 
 encoded using the Road Network Embedding (RNE) technique. The distances between two points are 
calculated by subtracting their corresponding RNE encoding vectors and then computing the 1-norm 
(i.e., the maximum of the absolute values) of the resulting 
vector. The driver with the smallest distance is chosen to offer the ride. For increased efficiency, 
each coordinate of the RNE 
encoding is decomposed into blocks of bits of size $l = 1$, $l = 2$, or $l = 4$. 
These individual blocks (along with the corresponding position weights) are encrypted using techniques 
similar to that of property-preserving hashing that employs bilinear pairings. 

The authors implement a 
prototype of the proposed protocol and evaluate it with real-world datasets to demonstrate its performance 
in practice. Further, all the riders and drivers are assumed to share two secret keys and all the entities
are trusted to not reveal these keys to others. The ORH SP is assumed to be \textit{honest-but-curious}, 
i.e., it follows the protocol correctly but may have intentions to learn about riders' and drivers' locations by         
monitoring the encrypted RNE encodings and the decrypted distances that it obtains as part of the protocol for ride matching. Also, the SP is not permitted to collude with the riders and drivers. The 
authors claim from their security analysis in \cite[Section V.B]{XieGJ21} that their protocol offers
strong privacy guarantees for riders' and drivers' RNE coordinates from the eyes of the ORH SP.  

\noindent\textbf{Our contribution}: we disprove the above mentioned privacy claim by exhibiting a straightforward attack
on the protocol from \cite{XieGJ21}. We show that the honest-but-curious ORH SP can completely recover the RNE coordinates (and hence 
the actual locations) of the rider and all the responding drivers in each and every ride request query. 
Hence, the primary security goal behind the design of the protocol from \cite{XieGJ21} is compromised. The key idea 
behind our attack is to make use of the decryptions of (signed) differences of individual blocks of the RNE 
coordinates of the drivers and the rider that the SP actually obtains after executing the protocol. Hence, our attack is 
robust against changes made to the ``encryption scheme'' deployed in the protocol so long as the SP still 
obtains the decryptions of partial distances.

\section{Recap of the ORH Protocol  \cite{XieGJ21}}
 Next, we briefly recap the technical details of the protocol from \cite{XieGJ21}, presenting only those details 
necessary to demonstrate our attack. In particular, we will \textit{not} describe the details of the 
encryption and decryption procedures.  

Any location $u$ is encoded using the RNE technique as $S(u) = (S_1(u), S_2(u), \ldots, S_\eta(u))$, where 
$\eta$ is a suitably chosen integer and each $S_i(u)$ is an unsigned integer in a given range. Let $R$ be 
the rider requesting a ride and $D_k$ be the $k$\textsuperscript{th} driver who responds to this ride 
request. Then the 
distance between $R$ and $D_k$ is computed as $\max_{i=1}^{\eta} |S_i(u_R) - S_i(u_{D_k})|$. The SP does 
the above distance computation in the clear (i.e., on plaintext values) in an indirect way (otherwise, 
the location encodings are revealed directly to the SP). The SP then selects the driver with the least 
distance to offer the ride (see \cite[Algorithms 1 -- 7]{XieGJ21}).

In order to reduce the number of ciphertexts, each coordinate $S_i(u_R)$ of the rider is further 
decomposed into $m$ many message blocks $v_j$ $(j=0,1,2,\ldots,m-1)$ of $l$ 
bits each ($l$ is typically 1 or 2 or 4). Hence, $S_i(u_R) = \sum_{j=0}^{m-1} v_j \cdot 2^{jl}$. Similarly, let 
$v^*_{k,j}$ $(j=0,1,2,\ldots,m-1)$ denote the blocks corresponding to the $i$\textsuperscript{th} coordinate $S_i(u_{D_k})$ of the 
$k$\textsuperscript{th} responding driver. The rider and the drivers encrypt each such block individually 
using a technique similar to property-preserving hashing that employs bilinear pairings. Actually, the 
rider sends to the SP the guessed scaled (signed) differences $(z_n - v_{j})\cdot 2^{jl}$ in the masked 
form (and in some randomly permuted order) for every value of $z_n$ in the range $(0,1,2,\ldots,2^l-1)$. 
Using some equality checks using pairings, 
the SP learns the actual difference  $(v^*_{k,j} - v_{j})\cdot 2^{jl}$ 
(and hence the value $v^*_{k,j} - v_{j}$) in the clear for each such block (see \cite[Algorithm 5]{XieGJ21}). The SP then
uses these intermediate values to compute the actual RNE distances (as plaintexts) between the rider and 
each responding driver.
The correctness of the above protocol is based on the below simple fact:
\begin{lemma}
\label{lem:zx}
Let $x \in \{0,1,2,\ldots,2^l-1\}$. Given all the (signed) differences $\{z-x\;|\; 0 \le z \le 2^l-1 \}$ (in any order), then we can uniquely recover $x$. 
\end{lemma}
Note that the computational cost for the rider increases exponentially w.r.t. the parameter $l$. Also, 
note that the differences have to be necessarily signed in order to correctly sum up the partial differences. 

\section{Our Attack}
At first look, it appears that given only the differences between the rider's and a driver's 
individual $l$-bit blocks, we may not be able to recover $S(u_R)$ nor $S(u_{D_i})$ as there
will be too many blocks to make an enumeration over the guesses feasible. 
Instead, our key idea is to observe, for a given block $v_j$, the differences $v^*_{k,j} - v_j$ of the corresponding 
blocks for many drivers (i.e., as $k$ varies over all the responding drivers). Typically, there will 
be several drivers who respond for a ride request (particularly so in densely populated areas). So if we can get
$2^l$ many drivers each having a different value for the corresponding block, then we can apply Lemma \ref{lem:zx} to
uniquely recover the rider block $v_j$. Next, using the given differences $v^*_{k,j} - v_j$, we can then recover the 
block $v^*_{k,j}$ for all the drivers, and eventually the original RNE encoding vectors and the actual locations themselves.

Now the question is what is the expected number of drivers who need to be present in order to ensure
that every possible $l$-bit value $v^*_{k,j}$ is covered? For the ease of analysis, we assume that these
blocks are uniform randomly and independently distributed. Then our problem is exactly the same as a 
classical problem in probability theory called the \textit{coupon collector's problem} \cite{coupon_collector}. 
We obtain that the expected
number of drivers required to be present is $2^l \cdot (1+\frac{1}{2}+\frac{1}{3}+\ldots+\frac{1}{2^l})$. The Table \ref{tab:exp}
lists the expected number of drivers needed to successfully mount our attack, and these numbers
are reasonable in the ORH context. Note that our attack is computationally very efficient.    

\begin{table}[h]
\centering
\caption{Expected no. of drivers needed to successfully mount our attack for different block sizes $l$.} 
\label{tab:exp} 
\begin{tabular}{ |c|c| } 
 \hline
 $l$ & Expected no. of drivers\\
 \hline\hline 
 1 & 3\\
 \hline 
 2 & 9\\ 
 \hline
 3 & 22\\ 
 \hline
 4 & 55\\
 \hline
\end{tabular}
\end{table}

\noindent\textbf{Example}: we now illustrate our attack for the setting $l=1$. Given the difference
$v^*_{k,j} - v_j$ for 1-bit blocks $v^*_{k,j}, v_j \in \{0,1\}$ (for all the drivers that are indexed 
by  $k$ ), we need to recover $v_j$ first and eventually $v^*_{k,j}$. There can only be four 
possibilities for the $(v^*_{k,j}, v_j)$ pair: $(0,0)$, $(0,1)$, $(1,0)$, $(1,1)$. These pairs yield the 
(signed) differences $0$, $-1$, $1$, $0$, respectively. Therefore, if the difference is $-1$ or $1$, then
we can uniquely recover the pair. If the difference is $0$, then we have two choices for the pairs: $(0,0)$ or $(1,1)$. In this case, we need to look for the differences of the corresponding blocks for a few more drivers to recover the pair uniquely. Like this every bit-block is reconstructed with no uncertainty and hence leading to the recovery of the actual RNE encoding vectors of the rider and all the drivers.

The authors of  \cite{XieGJ21} have also proposed an optimised version of the above protocol. Our attack applies to this case as well as the SP still has access to the decrypted partial distances like previously.

\section*{Acknowledgment}

This work was funded by the Infosys Foundation Career Development Chair Professorship grant for the author.

\bibliographystyle{IEEEtran}
\bibliography{refs}


\end{document}